\documentclass[12pt,a4paper]{article}
\usepackage[dvips]{epsfig}
\title{
  {\vspace{-3cm} \normalsize \hfill
    \parbox{38mm}{MS-TPI-00-3 \\
                  hep-th/0005084}  }\\[25mm]
  Distribution of Instanton Sizes in a Simplified Instanton Gas Model
  }
\author{Gernot M\"unster and Christel Kamp\\
        Institut f\"ur Theoretische Physik I,
        Universit\"at M\"unster\\
        Wilhelm-Klemm-Str.~9, D-48149 M\"unster, Germany\\
        e-mail: munsteg@uni-muenster.de}
\date{revised, July 14, 2000}
\begin{document}
\maketitle

\begin{abstract}
We investigate the distribution of instanton sizes in the framework of a
simplified model for ensembles of instantons.  This model takes into
account the non-diluteness of instantons.  The infrared problem for the
integration over instanton sizes is dealt with in a self-consistent
manner by approximating instanton interactions by a repulsive hard core
potential.  This leads to a dynamical suppression of large instantons.
The characteristic features of the instanton size distribution are
studied by means of analytic and Monte Carlo methods.  In one dimension
exact results can be derived.  In any dimension we find a power law
behaviour for small sizes, consistent with the semi-classical results.
At large instanton sizes the distribution decays exponentially.  The
results are compared with those from lattice simulations.
\end{abstract}
%
\section{Introduction}
%
%
\subsection{Instantons in gauge theories}

In non-abelian gauge theories different topologically nontrivial
configurations have been made responsible for non-perturbative features
\cite{tHooftConf,Mandelstam,Polyakov,tHooft79,Mack}.  An important class
are instantons, which are solutions of the Euclidean field equations
with non-vanishing topological charge \cite{Belavin}.  They give
contributions to the saddle-point approximation of Euclidean functional
integrals, which lead to non-perturbative effects \cite{tHooft76,CDG}.
For a review see \cite{Schaefer}.

In the dilute gas approximation \cite{CDG} one considers superpositions
of single instantons as quasi saddle points of the action.  These
configurations are characterized by the instanton positions
$\{\mathbf{a_j}\}$ in four-dimensional space-time, the instanton sizes
$\{\rho_j\}$, and other internal parameters.  For a single instanton the
action is
\begin{equation}
S_1 = \frac{8 \pi^2}{g_0^2} \,,
\end{equation}
where $g_0$ is the bare gauge coupling constant.  Taking into account
quadratic fluctuations around the one-instanton solution
\cite{tHooft76,CDG} its contribution to the functional integral is
\begin{equation}
Z_1 = \int\!\!d^4 \mathbf{a} \int\!\!d\rho \,
C \rho^{-5} \exp \left\{- \frac{8 \pi^2}{g^2(1/\rho)}\right\} \,,
\end{equation}
where $g(\mu)$ is the running coupling.  In the case of supersymmetric
Yang-Mills theory this formula has even been established at the
two-loop-level, in ordinary Yang-Mills theory there are higher-order
corrections to the integrand \cite{Morris}.  In one-loop order the
running coupling obeys
\begin{equation}
g^2(\mu) = \frac{8 \pi^2}{b \log (\mu / \Lambda)}
\end{equation}
where
\begin{equation}
b = \frac{11 N}{3} \,,
\end{equation}
and $\Lambda$ is the renormalization-group invariant scale parameter,
so that
\begin{equation}
\label{Z1}
Z_1 = \int\!\!d^4 \mathbf{a} \int\!\!d\rho \,
C \rho^{-5} (\rho \Lambda)^{b} \,.
\end{equation}
Therefore the integrand is proportional to $\rho^{b - 5}$ and increases
with increasing instanton size.

The space-time integral over the instanton position gives the usual
volume factor, which is needed in the large volume limit in order to get
an extensive free energy.  The integral over $\rho$, however, represents
an infrared problem.  Where the instanton density becomes important, for
$\rho \geq O(1/\Lambda)$, we leave the region of valididty of the
semiclassical expansion because the running coupling becomes too large.

It should be noted that the apparent infrared divergence in (\ref{Z1})
is an artifact of using the one-loop formula for $g^2(1/\rho)$ long
after it has become invalid, i.e.\ for $\rho \geq 1/\Lambda$.
Nevertheless we are confronted with the infrared problem in the size
integration for the single instanton contribution $Z_1$.  If the
semiclassical approximation is meaningful at all, a solution of this
problem in the context of the full instanton ensemble is required.

The quasi saddle points composed of any number of instantons and
anti-instantons are treated as independent in the dilute gas
approximation.  Consequently their contributions exponentiate in the
usual way.  The problem with the integration over the sizes persists and
has to be dealt with.  The simplest way is to cut the integrations off
at some ad-hoc value $\rho_c$.  But since the integrand increases with
increasing $\rho_j$ the dominant contribution comes from large $\rho_j$
near the cut-off where the assumption of diluteness fails.  Moreover the
introduction of an ad-hoc cut-off leads to inconsistencies with the
renormalization group \cite{IMP}.

In order to solve the problem of the instanton size integrations is has
been proposed that instanton sizes are cut off in a dynamical way
\cite{IMP,Mue82}.  The dynamical cut-off should originate from
configurations where instantons start to overlap.  Configurations of
overlapping instantons have an action which deviates from the sum of the
single instanton actions.  Therefore large instantons feel an
interaction.  Additionally, the fluctuations around the multi-instanton
configurations contribute to the instanton interaction.  The interaction
between instantons is expected to suppress overlapping instantons and to
result in a dynamical self-consistent cut-off.

Some consequences of this picture have been discussed in
\cite{IMP,Mue82}, based on certain assumptions about the repulsive
instanton interactions.  In \cite{Mue82} a temporary infrared cut-off
was introduced by means of a finite space-time volume $V$.  The large
$V$ limit was then considered with the help of renormalization group
arguments.  This led to some general results independent of the specific
form of the repulsive instanton interactions.  In particular, a finite
renormalization factor
\begin{equation}
\frac{4}{b} = \frac{12}{11 N}
\end{equation}
appears in some quantities, e.g.\ in the trace anomaly \cite{IMP},
correcting inconsistencies with the renormalization group. The same
factor is conjectured to multiply the instanton singularities in the
Borel plane, which then coincide with the infrared renormalons. The
conclusions were supported by considering a model of the instanton
ensemble, where the repulsive interaction is approximated by a hard
core.

The theory of instanton ensembles with a dynamical size cut-off has been
developed further by Shuryak \cite{Shuryak82}, see \cite{Schaefer} for a
review.  In his model of an ``instanton liquid'' various observables
have been calculated and related to hadronic phenomenology.

In connection with the dynamical cut-off the distribution of instanton
sizes is of central importance.  The size-distribution reflects the way
in which large instantons are suppressed and thus gives information
about the instanton interactions.  In recent years it has been studied
by means of lattice Monte Carlo calculations by different groups
\cite{FGPS96,FGPS97,Smith,Hasenfratz}.  For small sizes the distribution
is predicted to be
\begin{equation}
\label{smallrho}
n(\rho) \sim \rho^{b-5}
\end{equation}
by the dilute gas approximation as well as by the ``instanton liquid
model'', in accordance with Eq.~(\ref{Z1}).  For large sizes $\rho$,
where the dynamical cut-off is in effect, not much is known about the
distribution.  There are arguments \cite{IMP,Dyakonov,Shuryak99} in
favour of a suppression like
\begin{equation}
\label{largerho}
n(\rho) \sim \exp (-c \rho^p) \qquad \mbox{with} \quad p = 2 \,.
\end{equation}

In this article we investigate the distribution of instanton sizes in a
model \cite{Mue82} where the instanton interactions are approximated by
a repulsive hard core of variable size.  Although this approximation
appears to be crude, the general features of the instanton ensemble with
a dynamical cut-off are present.  In particular, using analytical and
numerical methods we calculate the asymptotic behaviour for small and
for large sizes $\rho$ and compare them with results from Monte Carlo
simulations of lattice gauge theory.  More details can be found in
\cite{Kamp}.

In \cite{Mue82} it has been conjectured that the distribution $n(\rho)$
is affected by the finite renormalization factor $4 / b$ in such a way
that for small $\rho$ asymptotically
\begin{equation}
\label{wrong}
n(\rho) \sim \rho^{-1} \rho^{\frac{4}{b}(b-4)} \,.
\end{equation}
Using the simplified model, we shall show below that this conjecture is
wrong and that instead the semiclassical result (\ref{smallrho}) holds.
%
%
\subsection{Simplified model for ensembles of instantons}

Consider an ensemble of instantons in $d$ space-time dimensions.  In the
spirit of \cite{Mue82} we introduce a finite volume $V$ and study the
approach to the thermodynamic limit $V \rightarrow \infty$.

In the sector with instanton number $K$ the partition function is
written as
\begin{equation}
Z_K(V) = \frac{C^K}{K!} \int \prod_{i=1}^K d{\bf a}_i d\rho_i
\prod_{j=1}^K \rho_j^{b-d-1}
\mathrm{e}^{-U(\{{\bf a}_k\},\{\rho_k\})} \,,
\end{equation}
where the instanton positions and radii are denoted $\{{\bf a}_i,
\rho_i\}$. $C$ is a constant, whose numerical value is unimportant here,
$b = 11 N / 3$ for SU($N$) Yang-Mills theory, and $U$ represents the
interaction potential between instantons. Distances are measured in
units of $\Lambda^{-1}$.

In our simplified model the repulsive potential is approximated by a
hard core potential. The radius of an instanton core varies
proportional to the size $\rho$ of the instanton. In a finite volume $V$
this means
\begin{equation}
\mathrm{e}^{-U(\{{\bf a}_i\},\{\rho_i\})} =
\Theta(\{{\bf a}_i\},\{\rho_i\})
\end{equation}
with
\begin{displaymath}
\begin{array}{ll}
\Theta(\{{\bf a}_i\},\{\rho_i\}) &= 1\ , \qquad \mbox{if}\ \left\{
\begin{array}{l}
1. \quad \| {\bf a}_i - {\bf a}_j \| >
\left( \frac{\tau}{v_1} \right)^{\frac{1}{d}}
({\rho}_{i} + {\rho}_{j}) \quad \forall i,j, \ \mbox{and}\\
2. \quad | {a}^{\mu}_i | < \frac{1}{2} V^{\frac{1}{d}} \quad
\forall \mu, i, \ \mbox{and}\\
3. \quad 0 < \rho_i <
\frac{1}{2} \left( \frac{v_1}{\tau} V \right)^{\frac{1}{d}}
\quad \forall i
\end{array}
\right. \\
\Theta(\{{\bf a}_i\},\{\rho_i\}) &= 0\ , \qquad \mbox{else} .
\end{array}
\end{displaymath}
Here
\begin{equation}
v_1=\frac{{\pi}^{\frac{d}{2}}}{\Gamma\left(\frac{d}{2}+1\right)}
\end{equation}
is the volume of the unit sphere in $d$ dimensions. The parameter $\tau$
specifies the effective volume $\tau \rho_j^d$ of an instanton and is of
the order of $v_1$.

We introduce a reduced distribution by
\begin{equation}
Z^{red}_K(V,\rho_K) = \frac{C^K}{K!} \int \prod_{i=1}^{K} d{\bf a}_i
\prod_{j=1}^{K-1} d\rho_j \prod_{k=1}^{K} \rho_k^{b-d-1}
\Theta(\{{\bf a}_l\},\{\rho_l\}) \,,
\end{equation}
such that
\begin{equation}
Z_K(V) = \int\!\!d\rho \ Z^{red}_K(V,\rho) \,.
\end{equation}

For the total system with variable instanton number the grand canonical
partition function is
\begin{equation}
Z(V) = \sum^{\infty}_{K=0} Z_K(V), \qquad \mbox{where} \quad Z_0(V)=1\,.
\end{equation}
We do not distinguish between instantons and anti-instantons in this
model.  In this way we neglect aspects of the interactions which differ
between instantons and anti-instantons, but we do not expect that they
play a significant role for our considerations.

The probability distribution of instanton numbers is given by
\begin{equation}
{\cal P}_K(V)=\frac{Z_K(V)}{Z(V)} \,.
\end{equation}

In order to define the probability distribution of instanton sizes one
has to specify how the sizes are sampled.  The definition should be made
in such a way that it is compatible with the Monte Carlo calculations to
be discussed later.  In the Monte Carlo runs configurations with a
variable number of instantons are produced.  A configuration of $K$
instantons contributes $K$ entries to the total histogram of instanton
sizes.  Therefore it has a relative weight proportional to $K$.
Correspondingly the probability distribution in the $K$-instanton sector
is normalized to $K$ \cite{Hill}:
\begin{equation}
\bar{n}_K(V,\rho) = K \frac{Z^{red}_K(V,\rho)}{Z_K(V)} \,.
\end{equation}
In the total ensemble the sizes are then distributed according to
\begin{eqnarray}
\bar{n}(V,\rho)&=&\sum^{\infty}_{K=1} {\cal P}_K(V) \bar{n}_K(V,\rho)\\
 &=&\frac{1}{Z(V)}\sum^{\infty}_{K=1} K Z^{red}_K(V,\rho)
\end{eqnarray}
with
\begin{equation}
\int_0^{\infty}\!\!\bar{n}(V,\rho) d\rho = \langle K \rangle_{V} \,.
\end{equation}
As the expectation value of the instanton number grows linearly with the
volume $V$ one is interested in the rescaled distribution
\begin{equation}
n(V,\rho) = \frac{\bar{n}(V,\rho)}{V} \,.
\end{equation}
In the following sections we study the properties of $n(V,\rho)$, and
its thermodynamic limit $n(\rho)$, respectively, utilizing analytical
approaches as well as Monte Carlo-methods.
%
%
\section{The one-dimensional instanton gas}

In $d=1$ dimensions the model can be solved exactly in the thermodynamic
limit.  This case illustrates some general features and can serve as a
testing ground for approximations used in higher dimensions.  Therefore
we shall discuss these results before we turn to the consideration of
other dimensions.

In the one-dimensional case the canonical partition function for a
system of spatial length $L$ can be written as
\begin{equation}
Z_K(L) = \frac{C^K}{K!}\int \prod_{i=1}^K da_i d\rho_i
\prod_{j=1}^K \rho_j^{b-2} \Theta(\{a_k\},\{\rho_k\})
\end{equation}
with
\begin{displaymath}
\begin{array}{ll}
\Theta(\{a_i\},\{\rho_i\}) &= 1\ , \qquad \mbox{if}\ \left\{
\begin{array}{l}
    1.) \quad | a_i - a_j | > \left(\frac{\tau}{2}\right)
    (\rho_{i}+\rho_{j}) \quad \forall i,j, \ \mbox{and}\\
    2.) \quad | a_i | < \frac{1}{2} L \quad \forall i, \ \mbox{and}\\
    3.) \quad 0 < \rho_i < \frac{L}{\tau} \quad \forall i
\end{array}
\right. \\
\Theta(\{a_i\},\{\rho_i\}) &= 0\ , \qquad \mbox{else} .
\end{array}
\end{displaymath}
This represents a system of rods with variable lengths on a line. A
pictorial representation is given in Fig.~\ref{stab}.
\begin{center}
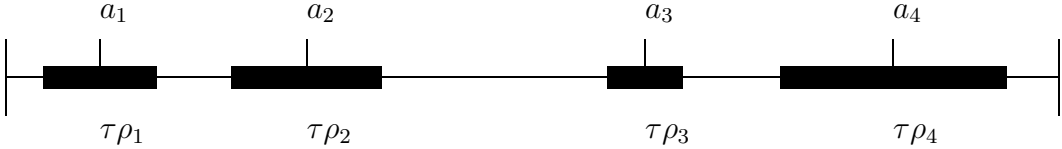
\begin{figure}[h]
\setlength{\unitlength}{1cm}
\begin{picture}(15,3)
\put (0.5,0.3){\line(0,1){1}}
\put (0.5,0.8){\line(1,0){14}}
\put (14.5, 0.3){\line(0,1){1}}
\put (1.75,0.8){\line(0,1){0.5}}
\put (1.75, 1.6){${a}_1$}
\put (1.75, 0){$\tau\rho_1$}
\put (4.5,0.8){\line(0,1){0.5}}
\put (4.5,1.6){${a}_2$}
\put (4.5,0){$\tau\rho_2$}
\put (9,0.8){\line(0,1){0.5}}
\put (9,1.6){${a}_3$}
\put (9,0){$\tau\rho_3$}
\put (12.3,0.8){\line(0,1){0.5}}
\put (12.3,1.6){${a}_4$}
\put (12.3,0){$\tau\rho_4$}
\linethickness{3mm}
\put (1,0.8){\line(1,0){1.5}}
\put (3.5,0.8){\line(1,0){2}}
\put (8.5,0.8){\line(1,0){1}}
\put (10.8,0.8){\line(1,0){3}}
\end{picture}
\caption{\rm \label{stab} One-dimensional instanton gas, $K=4$}
\end{figure}
\end{center}

Integration over the instanton positions $\{a_i\}$ yields the effective
free volume of $K$ indistinguishable particles on the line $L$:
\begin{equation}
\int \prod_{i=1}^K da_i\ \Theta(\{a_k\},\{\rho_k\}) =
\bar{\Theta}(\{\rho_k\})
\left( L - \tau \sum_{j=1}^{K} \rho_j \right)^K ,
\end{equation}
with
\begin{equation}
\bar{\Theta}(\{\rho_k\}) =
\left\{
\begin{array}{ll}
1, & \mbox{if} \quad \tau \sum^K_{j=1} \rho_j \le L \\
0, & \mbox{else},
\end{array}
\right.
\end{equation}
as can be shown by induction. With this result the partition function
reads
\begin{equation}
Z_K(L) = \frac{C^K}{K!} \int_0^{\frac{L}{\tau}} d\rho_K \cdots
\int_{0}^{\frac{L}{\tau}-\sum_{j=2}^{K}\rho_j}d\rho_1 \prod_{j=1}^{K}
\rho^{b - 2}_j \, \left(L - \tau \sum_{l=1}^K\rho_l\right)^K \,.
\end{equation}
The integrations over the $\rho_j$ can be carried out successively
employing
\begin{equation}
\int_0^1 \! x^{a} (1-x)^{b} dx =
\frac{\Gamma(a + 1)\Gamma(b + 1)}{\Gamma(a + b + 2)} \,,
\end{equation}
and one obtains for the size distribution
\begin{equation}
n(L,\rho) =
\frac{\sum_{K=1}^{\infty} K C \gamma^{K-1} (\Gamma(b(K-1)+2))^{-1}
\rho^{b - 2} (L - \tau \rho)^{b (K - 1) + 1}}
{\sum_{K=0}^{\infty} \gamma^{K} (\Gamma(b K + 1))^{-1} L^{b K + 1}}
\end{equation}
with
\begin{equation}
\gamma = \frac{C \Gamma(b - 1)}{\tau^{b - 1}} \,.
\end{equation}
The asymptotic behaviour for small instanton sizes is given by the power
law
\begin{equation}
n(L,\rho) \sim \mbox{const.}\, \rho^{b - 2} \qquad \mbox{for} \quad
\rho \rightarrow 0
\end{equation}
in the sense of
\begin{equation}
b - 2 = \lim_{\rho \to 0} \frac{\ln(n(L,\rho))}{\ln(\rho)} \,.
\end{equation}
In order to discuss the behaviour for large $\rho$ we take the infinite
volume limit
\begin{equation}
n(L,\rho) \stackrel{L \rightarrow \infty}{\longrightarrow} n(\rho) \,.
\end{equation}
The grand canonical sums can be evaluated by replacing them by integrals
over $K$ and performing a saddle point approximation, which becomes
exact in the large-$L$ limit. The result is
\begin{equation}
\label{nrho1}
n(\rho) = \frac{C}{b} \, \rho^{b - 2} \, \mathrm{e}^{- c \rho} \,,
\end{equation}
with
\begin{equation}
c = (C \Gamma(b - 2) \tau)^{\frac{1}{b}} \,.
\end{equation}
In addition to the power law with exponent $b - 2 = b - d - 1$ we
recognize an exponential suppression of large instanton sizes. The
exponent $p$ in the exponential, cp.\ Eq.~(\ref{largerho}), is equal to
1 in one dimension. Our next aim is to see how these results generalize
to higher dimensions $d$.
%
%
\section{The general instanton gas}

In higher dimensions, $d>1$, it is not possible to derive closed
expressions for the partition functions or the size distributions.  The
main difficulty is that the integrations over the instanton positions
and the radii cannot be decoupled.  In particular we have to use
approximations for the effective free volume of sets of instantons in
higher dimensions.  Nevertheless one can obtain approximate expressions
for $n(V,\rho)$ and derive its asymptotic behaviour for small $\rho$.

The canonical $K$-instanton partition function $Z_K(V)$ is written as
\begin{equation}
Z_K(V) = \frac{C^K}{K!} \int_{0}^{\bar V}\!\!dv
\int \prod_{i=1}^{K} d{\bf a}_i d\rho_i
\prod_{j=1}^{K} \rho_j^{\alpha} \
\Theta(\{{\bf a}_k\},\{\rho_k\})\,\delta(v-\tau\sum_l\rho_{l}^{d}) \,,
\end{equation}
where
\begin{equation}
\alpha = b - d - 1 \,,
\end{equation}
and
\begin{equation}
v = \tau \sum_l \rho_{l}^{d} \le {\bar V}
\end{equation}
is the total effective volume of instantons. The maximal accessible
volume ${\bar V} \le V$ accounts for the fact that spheres in dimensions
$d>1$ cannot completely occupy a given volume $V$.

An approximative decoupling of the integrations can be achieved through
the observation that for given $v = \tau \sum_j \rho_{j}^{d}$ the
product
\begin{displaymath}
\prod_{j=1}^{K} \rho_j^{\alpha}
\end{displaymath}
develops a sharp maximum at
\begin{equation}
\rho_j = \left( \frac{v}{K \tau} \right)^{1/d} = \rho_0 \,, \qquad
j \in \{1,...,K\}
\end{equation}
in the thermodynamic limit. The main idea is then to perform a saddle
point approximation for the $\rho$-integrations near this sharp maximum.
The integrations over the positions ${\bf a}_j$ then correspond to a gas
of hard spheres with equal radii $\rho_0$. With
\begin{equation}
\rho_j = \rho_0 + \delta_j
\end{equation}
and
\begin{equation}
\delta (v - \tau \sum_{j=1}^{K} \rho_{j}^{d}) \approx
\delta ( - \tau d \rho_0^{d-1} \sum_{j=1}^{K} \delta_j)
 = \frac{\rho_0^{1-d}}{2 \pi \tau d}
\int\!\!dq \ \mathrm{e}^{iq \sum_{j=1}^{K} \delta_j}
\end{equation}
the $\rho$-integrations can be solved straightforwardly.

In a similar way the reduced partition function $Z_K^{red}(V,\rho)$ can
be evaluated.  For a given radius $\rho$ of the $K^{th}$ instanton one
assigns an effective volume ${\bar V} - \tau \rho^{d}$ to the remaining
$K-1$ instantons and performs the saddle point approximation for the
integrations over $\rho_1, \ldots, \rho_{K-1}$ in terms of Gaussian
integrals.

The integral over the instanton positions
\begin{equation}
\frac{1}{K!} \int \prod_{j=1}^{K} d{\bf a}_j\
\Theta(\{{\bf a}_j\},\{\rho_0\})
\end{equation}
can be estimated with the help of geometrical considerations
\cite{Boltzmann,Hill,Reif,Mue82,Kamp}.
We use an approximation of the form
\begin{equation}
\frac{1}{K!} \int \prod_{j=1}^{K} d{\bf a}_j\
\Theta(\{{\bf a}_j\},\{\rho_0\})
\approx \frac{1}{K!} (V - v_{eff})^K \,,
\end{equation}
where
\begin{equation}
v_{eff} = h(d)\,v \,,
\end{equation}
and $h(d) = V / {\bar V}$ measures the inverse filling fraction of
spheres in a volume $V$. In the one-dimensional case we get $h(1)=1$
because a given length can be completely filled with rods. For $d>1$ we
consider $h(d)$ as a parameter. Lower and upper bounds are given by $1
\leq h(d) \leq 2^{d-1}$. For more details on this point see
\cite{Boltzmann,Hill,Reif,Mue82,Kamp}.

Using these expressions we get by some lengthy but straightforward
calculations for the canonical partition functions
\begin{equation}
Z_K(V) \approx \sqrt{\frac{\alpha}{2\pi}} \,\frac{\sqrt{K}}{d}
\left( \sqrt{\frac{2\pi}{\alpha}} \,
\frac{C}{(K \tau h(d))^{\beta-1}}\right)^K
\frac{\Gamma((\beta-1)K)}{\Gamma(\beta K+1)}\ V^{\beta K} \,,
\end{equation}
and for the reduced ones
\begin{displaymath}
Z_K^{red}(V,\rho) \approx \hspace{13cm}
\end{displaymath}
\begin{equation}
\sqrt{\frac{\alpha}{2\pi}} \,
\frac{\sqrt{K-1} \, C \, \rho^{\alpha}}{d}
\left(\sqrt{\frac{2\pi}{\alpha}} \,
\frac{C}{((K-1) \tau h(d))^{\beta-1}}\right)^{K-1}
\frac{\Gamma((\beta-1)(K-1))}{\Gamma(\beta(K-1)+2)} \
{\tilde V}^{\beta(K-1)+1}\,,
\end{equation}
where
\begin{equation}
{\tilde V} = V - h(d) \tau \rho^{d},\qquad
\beta = \frac{b}{d}\,.
\end{equation}

The next step is to perform the grand-canonical sums over the instanton
numbers $K$. As in the one-dimensional case, the sums can be evaluated
in the large volume limit by replacing them by integrals which are
calculated by means of the saddle point method. The error of this
approximation vanishes in the thermodynamic limit. For the size
distribution we get in this way
\begin{equation}\label{nrho}
n(\rho) = \frac{C d}{b} \, \rho^{b-d-1} \, \exp (- c \rho^{d}) \,,
\end{equation}
with
\begin{equation}
c^{\frac{b}{d}} = C \sqrt{\frac{2\pi}{b-d-1}}
\left( \frac{b}{d} - 1 \right)^{\frac{b}{d}-1}
\mathrm{e}^{-(\frac{b}{d}-1)} \ h(d) \, \tau \,.
\end{equation}
For $b \gg d$ this takes the form
\begin{equation}
c^{\frac{b}{d}} =
C \Gamma(\frac{b}{d} - 1) \, d^{-\frac{1}{2}} \ h(d) \, \tau \,,
\end{equation}
which agrees with the one-dimensional result.

The expression for $n(\rho)$ is consistent with the general expectation
mentioned in the introduction:  for small $\rho$ it grows powerlike with
an exponent $\alpha = b - d - 1$, and for large $\rho$ this power-law is
combined with an exponential decrease.

Although the canonical partition functions are dominated by
configurations where the instantons are densely packed, the exponent
$\alpha$ agrees with the one of the semiclassical dilute gas
approximation.  This result does not depend on the details of our
approximations and follows from the general structure of the occuring
terms in the grand-canonical sums.  The conjecture, made in
\cite{Mue82}, that due to the denseness of instantons the small-$\rho$
behaviour of $n(\rho)$ gets modified, is therefore wrong.

On the other hand, the value of the exponent $p = d$ in the exponential
decay at large $\rho$ should be considered with reservations, because it
depends on the saddle point approximations which have been made.  In
$d=1$ dimensions it is correct, but we would not be surprised, if in
higher dimensions the true value would differ from $d$.  In order to get
more insight into this question and to get an idea of the quality of the
approximations being made so far, we have also studied the instanton gas
by grand canonical Monte Carlo simulations.
%
%
\section{Monte Carlo simulations}

Usually Monte Carlo calculations are done in the canonical ensemble.  In
our case the particle number has to change and it is necessary to
simulate a grand canonical ensemble.  Simulations of grand canonical
systems are not very common.  They are rarely discussed in the
literature and some important details remain unclear.  Therefore it
appears appropriate to describe the algorithm we have used in our
calculations.  For related work on this topic we refer to
\cite{Adams74,Adams75,Barker,Valleau,Vanmegen,Sokal}.
%
%
\subsection{Grand canonical Monte Carlo algorithms}

A stochastic process, which is realized in a Monte Carlo simulation, is
specified by a transition matrix $W(X,Y)$, where $X$ and $Y$ denote
states of the system.  For the purpose of a simulation $W$ is usually
decomposed as a product of two factors:  $\omega(X,Y)$ represents a
proposal probability for a transition from $X$ to $Y$, and $a_{XY}$
denotes the corresponding acceptance probability.  In addition to
normalization and ergodicity one has to require stationarity, which is
often fulfilled by demanding the stronger {\sc Metropolis} condition of
detailed balance:
\begin{equation}
\label{accept}
a_{XY} =
\min \left( 1, \, \frac{\omega(Y,X) P(Y)}{\omega(X,Y) P(X)} \right)\,.
\end{equation}
Here $P$ is the probability distribution, which we want to generate as
the stationary distribution of the underlying stochastic process. In our
context it is given by
\begin{equation}
P_K(V; {\bf a}_1, \ldots, {\bf a}_K, \rho_1, \ldots, \rho_K) =
\frac{1}{Z(V)} \, \frac{C^K}{K!}
\prod_{j=1}^{K}\rho_j^{\alpha} \ \Theta(\{{\bf a}_j\},\{\rho_j\}) \,.
\end{equation}
The states $X$ and $Y$ are characterized by the instanton number $K$
combined with the set of coordinates $\{{\bf a}_j,\rho_j\}$.

In canonical algorithms $\omega(X,Y)$ is usually chosen to be symmetric
so that it is omitted in (\ref{accept}) without further comments.  This
is not possible in a grand canonical ensemble, where one has to consider
transitions that change the instanton number.  Independent of the choice
of $\omega(X,Y)$ there will be additional volume factors in $a_{XY}$ for
processes that do not conserve the instanton number.  This results from
the asymmetry in particle creation and destruction.  If an instanton is
created one has to specify a probability for the generation of its new
coordinates, On the other hand, in the process of removing an instanton
such a probability does not occur.  In our case, for the
space-coordinates as well as for the radii we choose a uniform
distribution within the allowed volume.

In the algorithm three different kinds of steps occur with equal
probability: creation, destruction and movement of an instanton. With
the shortcut notation
\begin{displaymath}
\star=\left\{
\begin{array}{l}
1. \quad \| {\bf a}_i - {\bf a}_j \| >
\left( \frac{\tau}{v_1} \right)^{\frac{1}{d}}
({\rho}_{i} + {\rho}_{j}) \quad \forall i,j, \ \mbox{and}\\
2. \quad | {a}^{\mu}_i | < \frac{1}{2} V^{\frac{1}{d}} \quad
\forall \mu, i, \ \mbox{and}\\
3. \quad 0 < \rho_i <
\frac{1}{2} \left( \frac{v_1}{\tau} V \right)^{\frac{1}{d}}
\quad \forall i
\end{array}
\right.
\end{displaymath}
we have chosen the following transition rules, where $x$ denotes a
random number between 0 and 1.
\begin{itemize}
\item{Creation:}
\newline
The creation of a new instanton with number $K+1$ and coordinates
$({\bf a}',\rho')$ is proposed, and
\begin{displaymath}
X \rightarrow \left\{
\begin{array}{l}
Y, \quad
\frac{C V (V v_1 / \tau)^{\frac{1}{d}}}{2(K+1)}\ {\rho'}^{\alpha}\geq x
\mbox{ and } \star \\
X, \quad
\frac{C V (V v_1 / \tau)^{\frac{1}{d}}}{2(K+1)}\ {\rho'}^{\alpha} < x
\mbox{ or not } \star.
\end{array}
\right.
\end{displaymath}
\item{Destruction:}
\newline
The destruction of an instanton with randomly chosen number $j$ is
proposed, and
\begin{displaymath}
X \rightarrow \left\{
\begin{array}{l}
Y, \quad
\frac{2K}{C V (V v_1 / \tau)^{\frac{1}{d}}} \, \rho_j^{-\alpha} \geq x\\
X, \quad
\frac{2K}{C V (V v_1 / \tau)^{\frac{1}{d}}} \, \rho_j^{-\alpha} < x\,.
\end{array}
\right.
\end{displaymath}
\item{Movement:}
\newline
A movement of a randomly chosen instanton in a volume element
$[-\delta_a,\delta_a]^{d}\times[-\delta_{\rho},\delta_{\rho}]$ around the
original coordinates is proposed, and
\begin{displaymath}
X\rightarrow\left\{
\begin{array}{l}
Y, \quad \left(\frac{\rho'_j}{\rho_j}\right)^{\alpha} \geq x
\mbox{ and } \star \\
X, \quad \left(\frac{\rho'_j}{\rho_j}\right)^{\alpha} < x
\mbox{ or not } \star\,.
\end{array}
\right.
\end{displaymath}
\end{itemize}
The simulations were started with the empty configuration ($K=0$).
Measuring was started after the instanton number reached saturation.
%
%
\subsection{Simulation results}

\begin{figure}[ht!]
\begin{center}
\epsfig{file=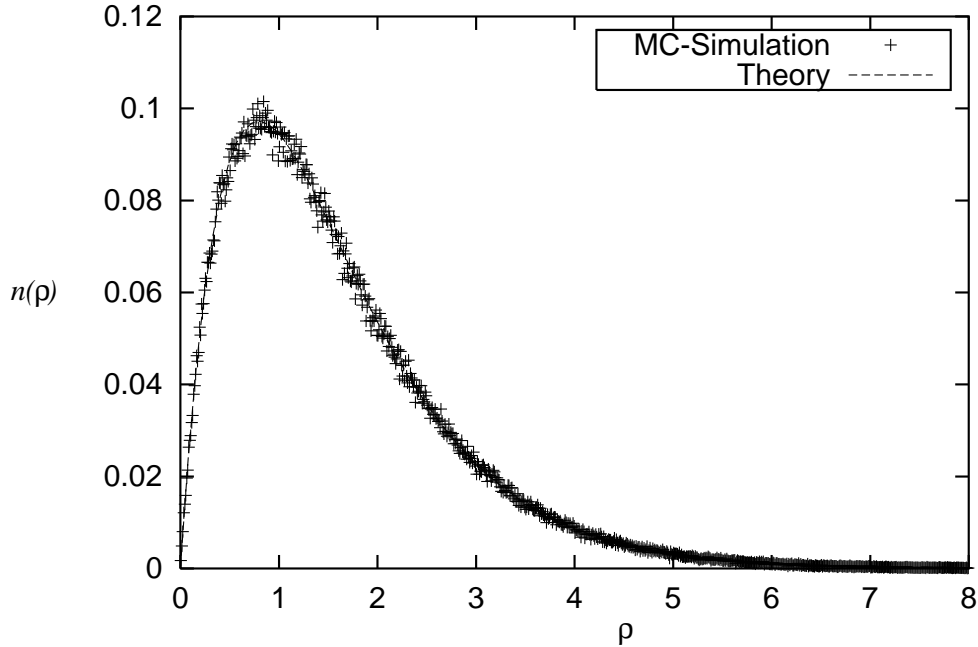,angle=270,width=13.4cm}
\caption{\rm \label{hist1} The size distribution $n(\rho)$ in $d=1$
from a Monte Carlo simulation with $C=1$, $\tau=2$, $\alpha=1$,
$V=L=2000$ in comparison with the predictions of formula (\ref{nrho1}).}
\end{center}
\end{figure}

In the case of $d=1$ dimensions the available exact result (\ref{nrho1})
provides a useful check of the Monte Carlo calculations.  In
Fig.~\ref{hist1} Monte Carlo data for $n(\rho)$ in $d=1$ are compared
with the exact formula.  The size $L$ has been chosen large enough such
that finite $L$ effects are negligible.  Obviously the Monte Carlo data
agree very well with the theoretical predictions, and the thermodynamic
limit has been approached sufficiently.

With this check on the Monte Carlo algorithm we proceed to the more
interesting case of four space-time dimensions ($d=4$).  In order to
compare the Monte Carlo data with the outcome of our analytical
approximations, Eq.~(\ref{nrho}), we have to make assumptions
concerning the parameter $h(d)=h(4)$ that describes the ability of
instantons to fill a given volume.  We consider three choices, namely
the lower bound $h_1=1$, the upper bound $h_2=2^{4-1}=8$ and their
geometric mean $h_g=2\sqrt{2}$.  The parameter $\tau$ is taken to be
equal to $v_1$. For the volume we chose $V=15^4$.  This is based on
simulations in different volumes, which showed that in this case the
thermodynamic limit was approximately reached within the errors of the
simulation. The parameter $\alpha$ is taken to be $\alpha = 7/3$, which
is the value for SU(2) gauge theory in 4 dimensions.

\begin{figure}[ht!]
\begin{center}
\epsfig{file=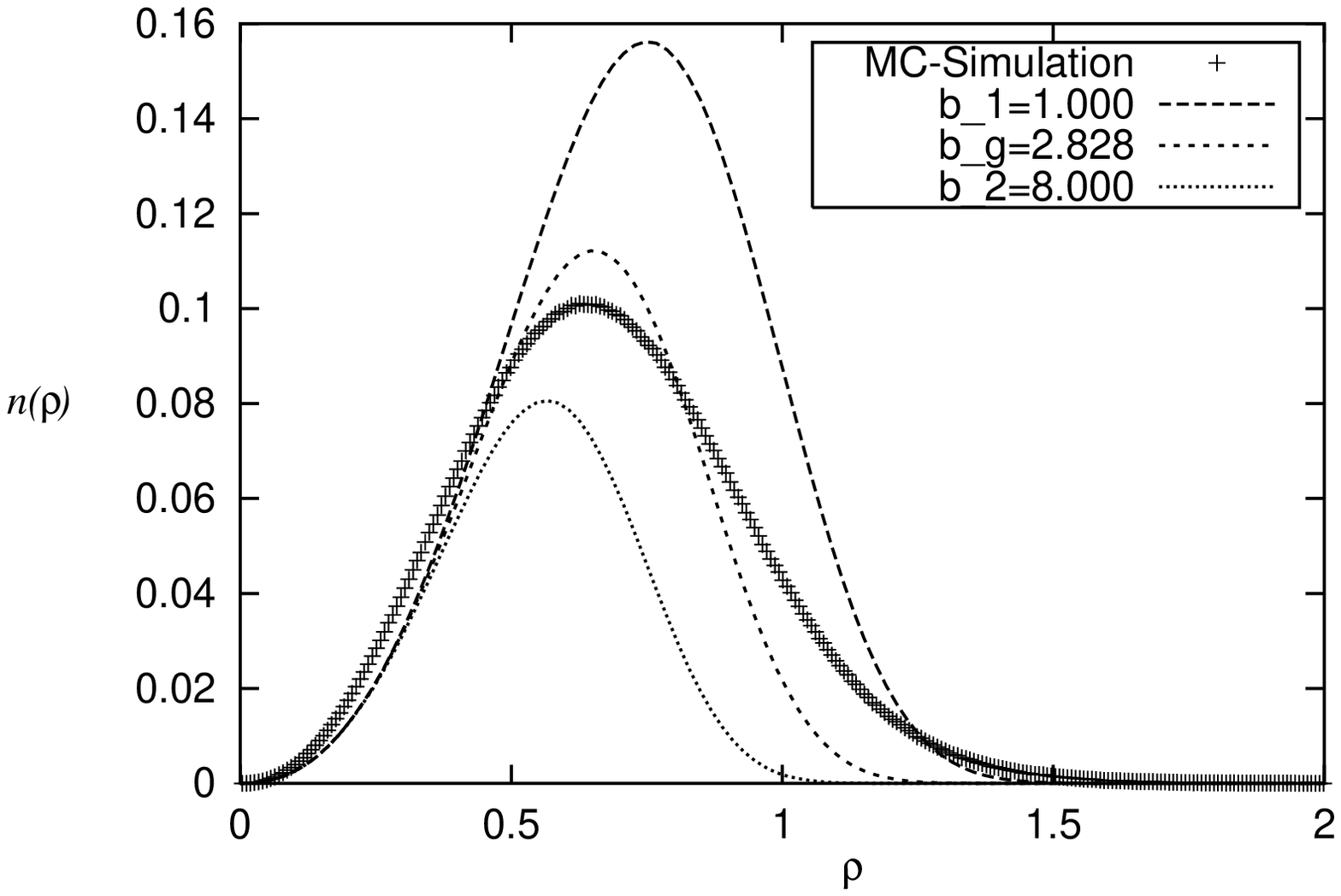,angle=270,width=13.4cm}
\caption{\rm \label{hist4} The size distribution $n(\rho)$ in $d=4$
dimensions from a Monte Carlo simulation with $C=1$, $\alpha=7/3$
(SU(2)), $V=15^4$ in comparison with the predictions of formula
(\ref{nrho}) for $h(4)=1$, $2 \protect\sqrt{2}$, and $8$.}
%
\epsfig{file=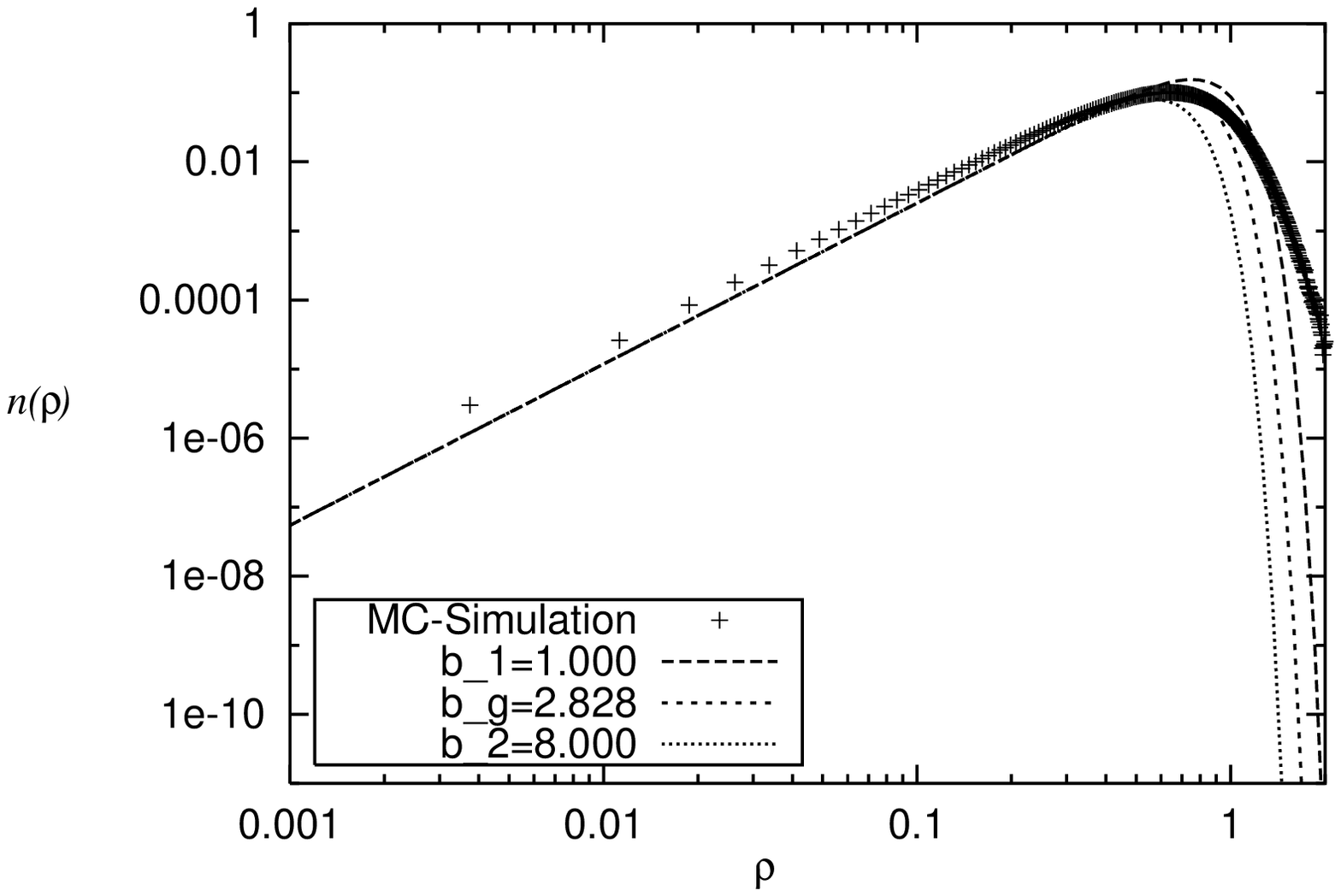,angle=270,width=13.4cm}
\caption{\rm \label{hist4ll} The size distribution $n(\rho)$ in $d=4$
dimensions from a Monte Carlo simulation with $C=1$, $\alpha=7/3$
(SU(2)), $V=15^4$ in comparison with the predictions of formula
(\ref{nrho}) for $h(4)=1$, $2 \protect\sqrt{2}$, and $8$, plotted on a
double logarithmic scale.}
\end{center}
\end{figure}

Fig.~\ref{hist4} shows the Monte Carlo data in comparison with the
analytical approximation. Near the maximum of the distribution the
approximation qualitatively reproduces the Monte Carlo results.
Furthermore, the growth of the distribution for small instanton radii
according to a power law with exponent $\alpha$ can be confirmed, as is
shown in Fig.~\ref{hist4ll}.

In order to study the behaviour of $n(\rho)$ for large $\rho$ we
considered the ratio
\begin{equation}
F(\rho) = \frac{n(\rho)}{\rho^{\alpha}} \,.
\end{equation}
Inspired by the theoretical results we tried fits of the form
\begin{equation}
F_{fit}(\rho) = a \exp (-c \rho^p) \,.
\end{equation}
The parameter $a$ was obtained by extrapolating $F(\rho)$ to small
$\rho$. The fit with parameters $c$ and $p$ was then obtained using the
Marquardt--Levenberg-algorithm. We performed fits for various choices of
the model parameters $C$ and $\alpha$. In agreement with the theoretical
results they showed that $c$ depends on $\alpha$, while $p$ is nearly
independent of it.

\begin{figure}[ht!]
\begin{center}
\epsfig{file=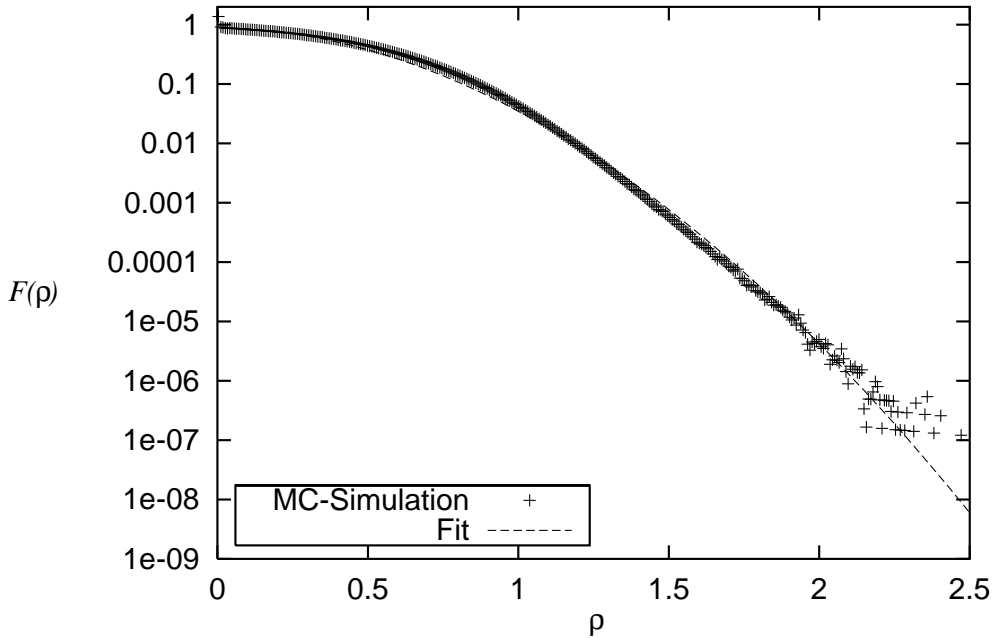,angle=270,width=13.4cm}
\caption{\rm \label{fit} $F(\rho) = n(\rho) / \rho^{\alpha}$ in $d=4$
dimensions from a Monte Carlo simulation with $C=1$, $\alpha=7/3$
(SU(2)), $V=15^4$ in comparison with the fit $F_{fit}(\rho)$ with
$a=0.89$, $c=3.24$, $p=1.92$, plotted on a logarithmic scale.}
\end{center}
\end{figure}

The main interest is in the exponent $p$.  We present the results for
the parameter set $C=1$, $\alpha=7/3$, $V=15^4$, because this value of
$\alpha$ is relevant for gauge theory with gauge group SU(2).  For
$\alpha=6$, the SU(3) case, the results for $p$ are the same within the
present errors.  We find $a\approx 0.89$, and the fit leads to $c=3.3
\pm 0.2$ and $p=1.9 \pm 0.2$.  In Fig.~\ref{fit} the result of a fit
in the interval $[0,2.25]$ is shown.

This Gaussian ($p=2$) decay of the probability distribution has
already been predicted by some authors under various assumptions
\cite{IMP,Dyakonov}. A recent approach based on an idea of dual
superconductivity \cite{Shuryak99} also leads to the prediction $p=2$.
Furthermore good agreement with the $SU(3)$ lattice gauge theory
calculations of Hasenfratz et al.\ \cite{Hasenfratz} was found.

The exponent $p=2$ differs from the one predicted by our approximate
analytical calculation, $p=d$.  The saddle point approximation being
made is, however, not beyond any doubt.  In that case the exponent
originates from the effective excluded volume being proportional to
$\rho^d$ at the considered saddle point.  This would also coincide with
the intuitive expectation based on the following picture.  In the
presence of a large instanton of size $\rho$ the remaining ones are
excluded from a volume $\sim \rho^d$.  If they behave like a dilute gas,
one would expect that the excluded volume yields a suppression factor
$\propto \exp (-c \rho^d)$.  The instanton ensemble in the effective
remaining volume is, however, dominated by dense configurations, as the
analytical calculation shows.  Therefore the intuitive picture should be
considered with reservations.  Indeed, the calculations of \cite{IMP}
take into account excluded volume effects in the framework of the theory
of grand canonical pair distribution functions and, also employing
certain approximations, arrive at $p=2$.

In recent years much effort has been devoted to lattice Monte Carlo
calculations of properties of the instanton ensemble.  There are still
ambiguities due to smoothing procedures and only data with little
statistics are yet available.  Nevertheless some quantitative statements
have been given.  Concerning the size distribution for small $\rho$,
lattice calculations appear to support the power law (\ref{smallrho})
rather than (\ref{wrong}).  A nice plot, using data of \cite{Smith}, can
be found in \cite{Ringwald}.  For the large-$\rho$ distribution, de
Forcrand et al.\ predict an exponential decrease with $p = 3 \pm 1$ from
their $SU(2)$ lattice data \cite{FGPS97}.  In contrast to this, Smith
and Teper conclude form their $SU(3)$ simulations a decay according to
$\rho^{-\xi}$ with $\xi \approx 10 \ldots 12$ \cite{Smith}.  %

We have studied the distribution of instanton sizes $\rho$ in the
framework of a model, where instanton interactions are approximated by a
hard core potential with variable radius.  This model incorporates the
basic features of a dynamical cut-off on large instanton sizes.  In the
one-dimensional case an exact formula can be derived, which yields a
power-like growth $\sim \rho^{\alpha}$ for small radii $\rho$ and an
exponential decay for large $\rho$.

In four space-time dimensions we employed analytical approximations as
well as Monte Carlo simulations.  The theoretical calculations
generalize the one-dimensional results and give a power-like behaviour
for small $\rho$.  For large radii $\rho$ they overestimate the decay
which is found in the Monte Carlo data.  Fits to the numerical Monte
Carlo results suggest a behaviour like
\begin{equation}
n(\rho) \stackrel{\rho\to\infty}{\sim} \exp (-c \rho^2) \,,
\end{equation}
in agreement with some other work on gauge theories.

The results indicate that our simplified model reproduces the main
features of instanton ensembles with a dynamical infrared cut-off.
Definite results about properties of instanton ensembles can of course
only be expected from future Monte Carlo calculations of lattice gauge
theories.

%

%

\begin{thebibliography}{99}

%
\bibitem{tHooftConf}
G.~'t Hooft,
in: {\it High Energy Physics}, Proc.\ European Phys.\ Soc.\ Int.\
Conf.\ 1975, ed.\ A.~Zichichi, Bologna 1976, p.\,1225.
%
\bibitem{Mandelstam}
S.~Mandelstam,
Phys.\ Reports \underline{23\,C} (1976) 245.
%
\bibitem{Polyakov}
A.~Polyakov,
Phys.\ Letters \underline{B\,59} (1975) 82.
%
\bibitem{tHooft79}
G.~'t Hooft,
Nucl.\ Phys.\ \underline{B\,153} (1979) 141.
%
\bibitem{Mack}
G.~Mack and V.~Petkova,
Ann.\ Phys.\ (N.Y.) \underline{123} (1979) 447;
Ann.\ Phys.\ (N.Y.) \underline{125} (1980) 117.
%
\bibitem{Belavin}
A.~Belavin, A.~Polyakov, A.~Schwartz and Y.~Tyupkin,
Phys.\ Letters \underline{B\,59} (1975) 85.
%
\bibitem{tHooft76}
G.~'t Hooft,
Phys.\ Rev.\ \underline{D\,14} (1976) 3432.
%
\bibitem{CDG}
C.~Callan, R.~Dashen and D.~Gross,
Phys.\ Rev.\ \underline{D\,17} (1978) 2717.
%
\bibitem{Morris}
T.~Morris, D.~Ross and C.~Sachrajda,
Nucl.\ Phys.\ \underline{B\,255} (1985) 115;
Phys.\ Letters \underline{B\,172} (1986) 40.
%
\bibitem{Schaefer}
T.~Sch\"afer and E.~Shuryak,
Rev.\ Mod.\ Phys.\ \underline{70} (1998) 323.
%
\bibitem{IMP}
E.-M.~Ilgenfritz and M.~M\"uller-Preu{\ss}ker,
Nucl.\ Phys.\ \underline{B\,184} (1981) 443.
%
\bibitem{Mue82}
G.~M\"unster,
Z.\ Phys.\ C, Particles and Fields \underline{12} (1982) 43.
%
\bibitem{Shuryak82}
E.~Shuryak,
Nucl.\ Phys.\ \underline{B\,198} (1982) 83.
%
\bibitem{FGPS96}
Ph.~de Forcrand, M.~Garc{\'\i}a P\'erez and I.-O.~Stamatescu,
Nucl.\ Phys.\ B (Proc.\ Suppl.) \underline{47} (1996) 777.
%
\bibitem{FGPS97}
Ph.~de Forcrand, M.~Garc{\'\i}a P\'erez and I.-O.~Stamatescu,
Nucl.\ Phys.\ \underline{B\,499} (1997) 409.
%
\bibitem{Smith}
D.A.~Smith and M.J.~Teper,
Phys.\ Rev.\ \underline{D\,58} (1998) 014505.
%
\bibitem{Hasenfratz}
A.~Hasenfratz and C.~Nieter,
Phys.\ Letters \underline{B\,439} (1998) 366.
%
\bibitem{Dyakonov}
D.I.~Dyakonov and V.Y.~Petrov,
Nucl.\ Phys.\ \underline{B\, 245} (1984) 259.
%
\bibitem{Shuryak99}
E.V.~Shuryak,
\texttt{hep-ph/9909458}.
%
\bibitem{Kamp}
C.~Kamp,
\textit{Untersuchung der Radiusverteilung in einem Instantongas mit
Hilfe analytischer Methoden sowie gro{\ss}kanonischer
Monte-Carlo-Verfahren},
Diploma thesis, University of M\"unster, 1999.
%
\bibitem{Hill}
T.L.~Hill,
\textit{Statistical Mechanics, Principles and selected Applications},
McGraw-Hill Book Company, 1956.
%
\bibitem{Boltzmann}
L.~Boltzmann,
\textit{\"Uber die Zustandsgleichung van der Waals},
in \textit{Wissenschaftliche Abhandlungen, Band III}, 1882-1995.
%
\bibitem{Reif}
F.~Reif,
\textit{Physikalische Statistik und Physik der W\"arme},
de Gruyter, 1976.
%
\bibitem{Adams74}
D.J.~Adams,
Mol.\ Phys.\ \underline{28} (1974) 1241.
%
\bibitem{Adams75}
D.J.~Adams,
Mol.\ Phys.\ \underline{29} (1975) 307.
%
\bibitem{Barker}
J.A.~Barker and D.W.~Henderson,
Rev.\ Mod.\ Phys.\ \underline{48} (1976) 587.
%
\bibitem{Valleau}
J.P.~Valleau and L.K.~Cohen,
J.\ Chem.\ Phys.\ \underline{72} (1980) 5935.
%
\bibitem{Vanmegen}
W.~van Megen and I.K.~Snook,
Mol.\ Phys.\ \underline{39} (1980) 1043.
%
\bibitem{Sokal}
A.D.~Sokal,
\textit {Monte Carlo Methods in Statistical Mechanics:
Foundations and New Algorithms},
Cours de Troisi\`eme Cycle de la Physique en Suisse Romande,
Lausanne, 1989.
%
\bibitem{Ringwald}
A.~Ringwald and F.~Schrempp,
Phys.\ Letters \underline{B\,459} (1999) 249.
%
\end{thebibliography}
\end{document}